# Spectrally Programmable Spin-Polarized Photocurrents in $WSe_2$-$NiPS_3$ Magnetic van der Waals Heterostructures

Rajesh Kumar Yadav[1], Michal Poplinger[1], Adi Levi[1], Adi Harchol[2], Nirman Chakraborty[2,3], Thomas Brumme[4], Thomas Heine[4,5,6*], Efrat Lifshitz[2*] and Doron Naveh[1*]

1. Faculty of Engineering, Institute for Nanotechnology and Advanced Materials, Bar-Ilan University, Ramat-Gan 52900, Israel.
2. Schulich Faculty of Chemistry, Solid State Institute, Russell Berrie Nanotechnology Institute, Helen Diller Quantum Information Center, Technion, Haifa 3200003, Israel
3. Eduard Zintl Institute of Inorganic and Physical Chemistry, Technical University of Darmstadt, 64287 Darmstadt, Germany
4. Faculty of Chemistry and Food Chemistry, Technische Universität Dresden, 01062 Dresden, Germany.
5. Center for Advanced Systems Understanding, CASUS, HZDR, 02826 Görlitz, Germany
6. Department of Chemistry and ibs for nanomedicine, Yonsei University, Seoul 120-749, Republic of Korea

[*] Corresponding authors: Doron.Naveh@biu.ac.il, ssefrat@technion.ac.il, thomas.heine@tu-dresden.de

Efficient generation and control of spin-polarized currents in semiconductors remain central challenges for spin-based electronics, particularly due to impedance mismatch and the reliance on magnetic fields or ferromagnetic contacts. Here, we introduce a materials platform for spectrally programmable spin transport based on a van der Waals heterostructure combining the antiferromagnetic semiconductor $NiPS_3$ with $WSe_2$. In a p–n diode architecture, circularly polarized excitation produces pronounced photoconductive resonances with spin polarization reaching 80% near the Néel temperature and persisting at ≈30% at room temperature. Remarkably, selected spectral bands retain their polarization sign across the magnetic phase transition, evidencing robust, spectrally protected spin-polarized current generation. Polarization-resolved photogalvanic measurements reveal a dominant circular injection-current mechanism, confirming spin-polarized carrier transport. First-principles calculations show that an applied electric field induces interfacial hybridization and spin-layer locking, giving rise to localized symmetry breaking and enhanced optical absorption while preserving global time-reversal symmetry. These results establish spectral tuning of excitation as a new control knob for spin transport, enabling spin-current generation without magnetic fields or polarization switching. Our findings position magnetic van der Waals heterostructures as a versatile platform for opto-spintronic functionality and spectrally programmable spintronic devices.

## 1. Introduction

Spintronics traditionally exploits the electron spin as an information carrier in solid-state systems, complementing or replacing charge-based electronics.[1–4] Opto-spintronics extends this rationale by using optical excitation and detection to control spin states.[1,5,6] Optical generation of spin-polarized carriers overcome a fundamental challenge of impedance mismatch in semiconductor spintronic devices.[5,7,8] Transition-metal dichalcogenides (TMDs) such as $WSe_2$ exhibit strong spin-orbit coupling (SOC) and broken inversion symmetry, leading to spin-valley coupling.[9–12] Circularly polarized light selectively addresses specific valleys, enabling optical control of coupled spin and valley degrees of freedom.[10] This has given rise to valley-opto-spintronics. The discovery of two-dimensional (2D) van der Waals (vdW) magnets has opened a new frontier for exploring low-dimensional quantum phenomena, giving rise to the emergence of optical generated spin physics.[13–15] Among them, $NiPS_3$ has emerged as a prototypical antiferromagnetic semiconductor, distinguished by its zigzag spin order, sharp excitonic resonances, and strong spin-lattice coupling.[16–20] Recent progress has demonstrated a remarkable versatility of $NiPS_3$ and its heterostructures for quantum photonics and spintronics.[7,21,22] Strain engineering creates localized chiral quantum emitters with near-unity circular polarization and single-photon purity through magnetic proximity coupling between $WSe_2$ excitons and local magnetic moments in $NiPS_3$.[22] Carrier doping can reversibly drive a magnetic phase transition from antiferromagnetic to ferrimagnetic, rendering the exchange interactions of the free carriers as dominant.[23]

Heterostructures of $NiPS_3$ with TMDs such as $WSe_2$ provide a rich arena for exploring magnetic proximity effects. When coupled to $WSe_2$, the antiferromagnetic order in $NiPS_3$ imparts chiral and valley-selective properties to excitons, in the absence of external magnetic fields, owing to local symmetry reduction resulted from strain fields.[22] Strain engineering at the heterointerface further enhances these effects, co-localizing excitons with localized ferromagnetic moments in $NiPS_3$ and amplifying valley polarization responses.[22] Moreover, integrating twisted bilayer $WSe_2$ moiré superlattices with $NiPS_3$ has revealed strain-sensitive amplification of moiré exciton valley polarization, pointing to a tunable interplay between excitonic texture and interfacial magnetism[24]. Together, these advances demonstrate that $WSe_2$-$NiPS_3$ heterostructures serve as versatile building blocks for hybrid quantum photonic devices, combining the optical robustness of TMDs with the magnetically correlated order of layered antiferromagnets. In this work, we focus on opto-spintronics in antiferromagnetic vdW heterostructures, demonstrating how optical excitation can be harnessed to robustly generate spin-polarized currents. Using a $WSe_2$-$NiPS_3$ p-n diode, we observe spectral resonances of the photocurrent circular dichroism. Polarization-resolved and temperature-dependent measurements reveal spectrally protected spin-polarized photocurrent bands that persist across the antiferromagnetic phase transition of $NiPS_3$. Combined photogalvanic experiments and first-principles calculations show that these effects originate from electric-field-induced interfacial hybridization and spin-layer locking, establishing spectral selectivity as a key control parameter for opto-spintronics.

## 2. Results and Discussion

### 2.1 $WSe_2$–$NiPS_3$ heterostructure p–n diode

Heterostructure devices of hBN-encapsulated p-type ~9 layers thick $WSe_2$ with ~100 nm thick n-type $NiPS_3$ (Figure 1a, b) constructs a diode, featuring a rectified current with an apparent temperature-dependent breakdown that complies with a phonon-assisted tunneling (Figure 1c).[25,26] The photocurrent at a strong reverse bias of -5V as a function of the exciting photon energy is displayed in Figure 1d: The current generated by left (blue line) and right (red line) circularly polarized light sums to the current generated with unpolarized (black line) light. Interestingly, the photocurrent spectra of the circularly polarized light feature complementary energy resonances.

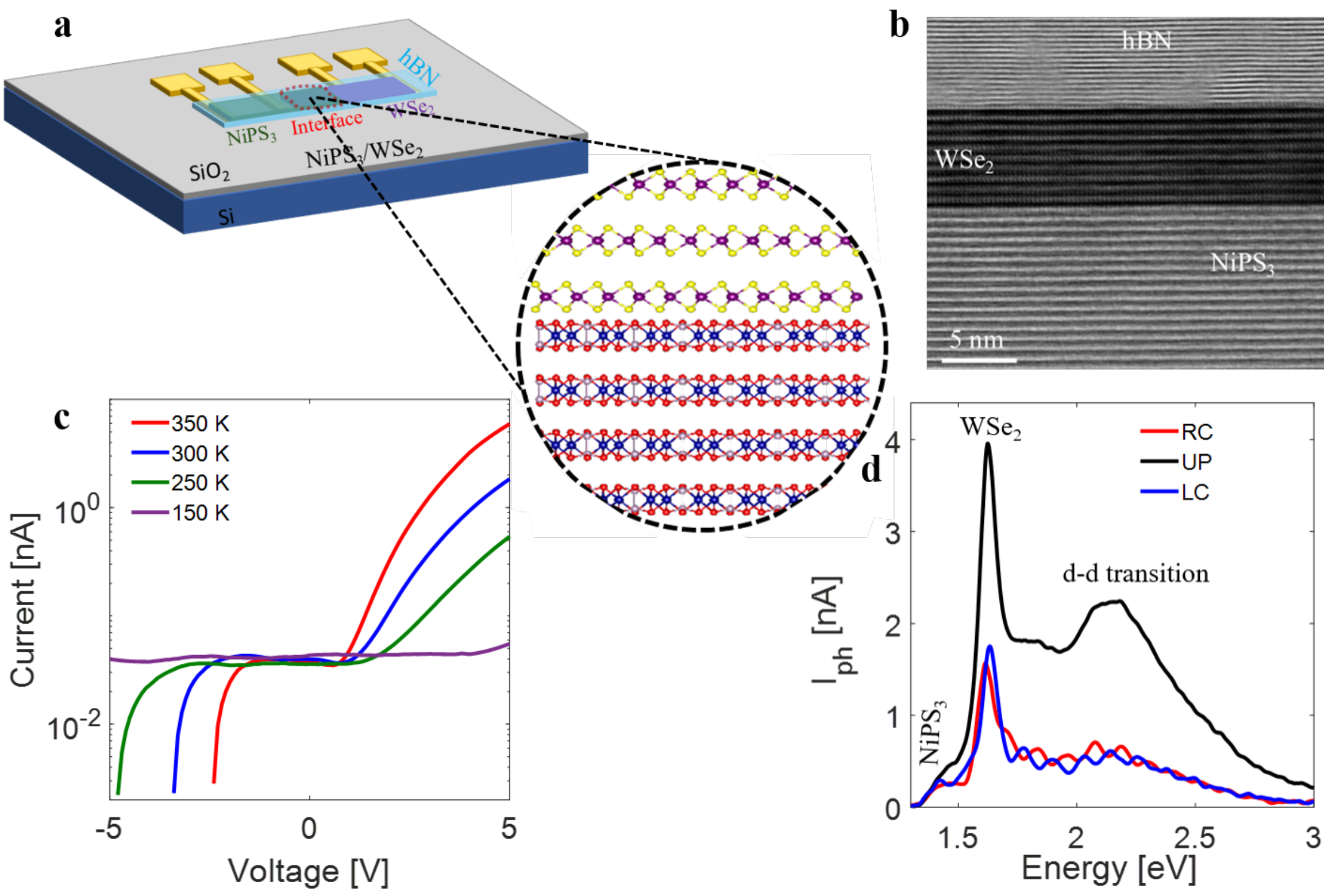

**Figure 1.** Heterostructure p-n diode of $WSe_2$ and $NiPS_3$. a) schematic description of the device, the top layer represents $WSe_2$ (W: purple; Se: yellow), while the bottom layer represents $NiPS_3$ (Ni: blue; S: red; P: gray). b) cross-sectional TEM micrograph of the interface, c), characteristic current-voltage curves of $WSe_2$-$NiPS_3$ diode at several temperatures, and d) the photocurrent response to randomly (black) polarized, right (red) and left (blue) circularly polarized light.

The onset of the photocurrent that appears as a weak response at ~1.45 eV, is associated with the absorption edge of $NiPS_3$.[18,21,27] Further to Figure 1d, a detailed study of the photocurrent spectral features, including their dependence on the applied voltage (Supporting Figure S2) and temperature (Figure 2), reveals further details. A prominent, sharp photocurrent peak appears at ~1.5 eV (Figure 2a), that is attributed to transitions involving $WSe_2$, featuring strong temperature dependence of the photocurrent, suggesting a phonon-assisted process. The blue shift of this peak at low temperature accords with the Varshney effect.[28] Around ~2 eV, there is a broad spectral feature that is associated with *d-d* transitions of $NiPS_3$.[29,30] The spectral features associated with d–d transitions (~2.3 eV) become more pronounced in devices with comparably thin (~10 nm) layers, likely due to the reduced conductivity of $NiPS_3$. Although thickness can affect the relative intensity of the spectral features, the interfacial selection rules remain robust, giving rise to circular dichroic resonances (Figure S11).

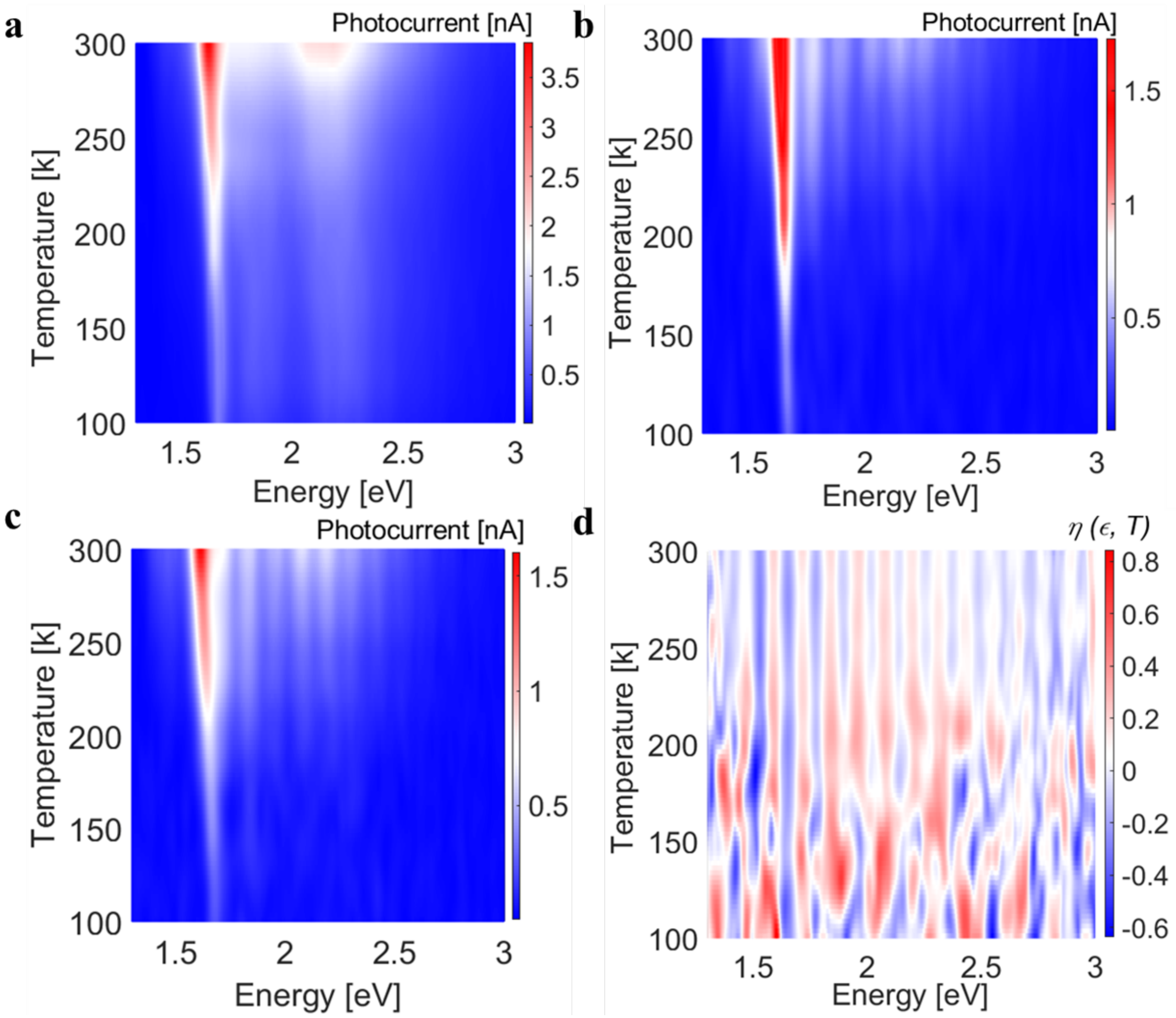

**Figure 2.** Temperature (100-300 K) dependent photocurrent spectra. a–c) Photocurrent spectrum as a function of temperature measured on the device with randomly polarized (a), left circularly polarized light (b), and right circularly polarized light (c). d) The resulting circular dichroic polarization spectrum.

## 2.2 Temperature-dependent and polarization-resolved photocurrent spectroscopy

The circular left and right polarizations appearing in Figure 2b, c show spectrally quantized photoconductive bands that superimpose to construct the unpolarized measurement of Figure 2a. The photoconductive spectral resonant bands suggest that an optical injection of spin polarized currents may be generated. The degree of photocurrent polarization per temperature and photon energy is defined as -

$$\eta(\varepsilon,T)=\frac{I_{ph}^{R}(\varepsilon,T)-I_{ph}^{L}(\varepsilon,T)}{I_{ph}^{R}(\varepsilon,T)+I_{ph}^{L}(\varepsilon,T)} \qquad (1)$$

where $I_{ph}^{x}(\varepsilon,T)$ is the photocurrent generated with photons of energy $\varepsilon$ and $x$={$R, L$} is {right, left} circular polarization at temperature $T$, as constructed in Fig. 2d from the data of Fig. 2b-2c. Here the photocurrent reaches a polarization of ~80% close to the Néel temperature, and ~30% at room temperature. Some degree of disorder in the polarization of photoconductive resonances (Figure 2d) appears near the Néel temperature of $NiPS_3$ (150K). Interestingly, some protected photoconductive bands can be observed in the spectrum, remaining negative (blue) or positive (red) at any temperature. One such photoconductive band appears for photons energy of ~2 eV, for which a monochromatic photogalvanic measurement sheds light on the underlying mechanism of the observation on monotonic optically pumped spin polarized currents. The energy of the excitation photon serves as an independent control parameter for spin-selective photocurrent generation. Through the tailored construction of heterostructures, interfacial hybridization and the lifting of degeneracies can drive a variety of optical transitions that yield finite spin injection.

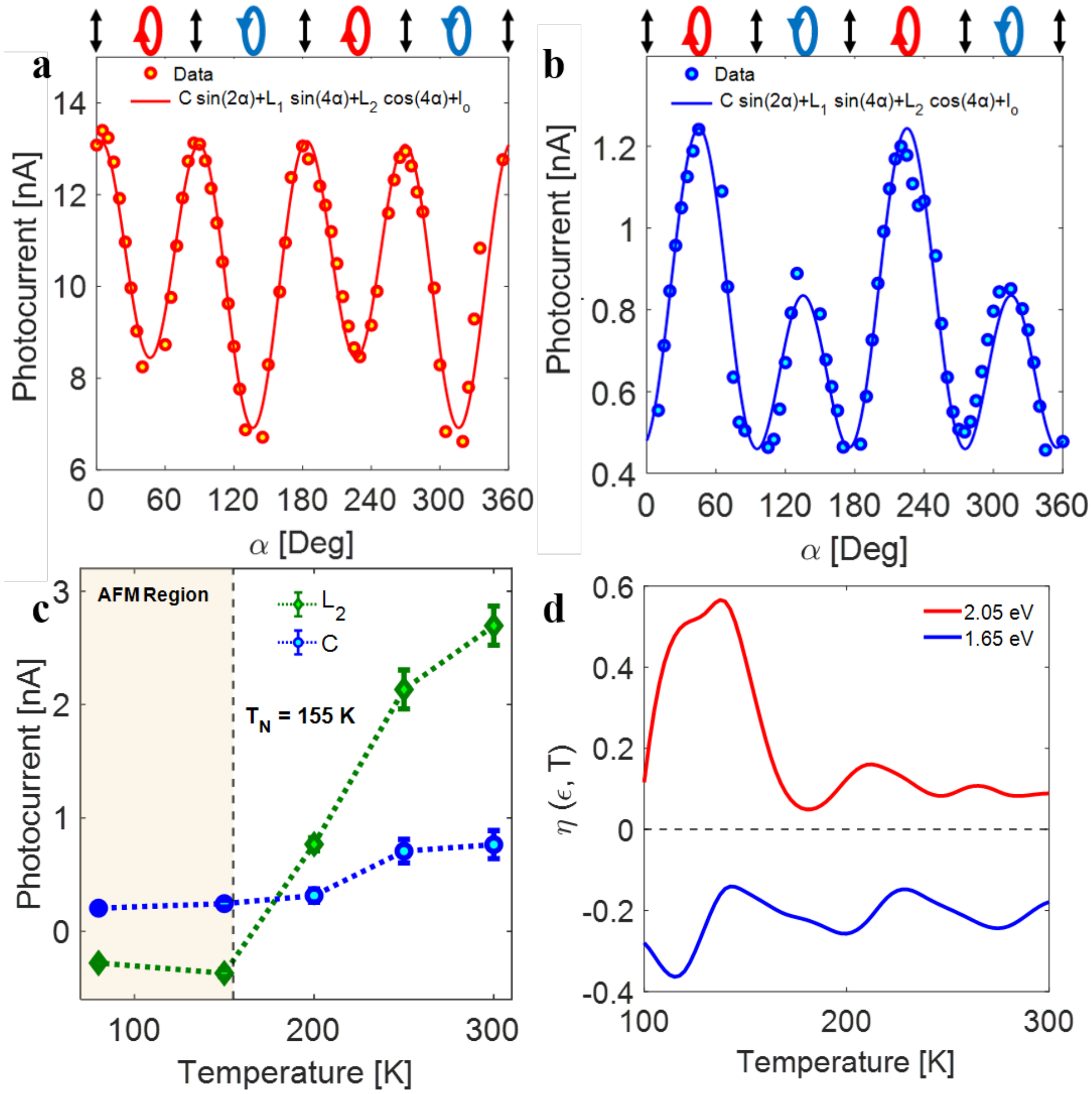

**Figure 3.** Temperature (80-300 K) dependent Photogalvanic response at 45º oblique incidence. a, b) Photocurrent (data: circles; fit: line) collected at 300K (a) or 80K (b), excited with 1.96 eV laser and resolved by the angle $\alpha$ between the polarization of the laser and the axis of a $\lambda/4$ retarder. c) The amplitude of the circular (blue circles) and linear (green diamonds) polarization components of the photocurrents as function of temperature, and d) Temperature-dependent photocurrent polarization $\eta(\varepsilon, T)$ at photon energies of 2.05 eV (red) and 1.65 eV (blue).

**Figure 3** displays the monochromatic photogalvanic measurement with a He-Ne laser (1.96 eV) at 45° angle of oblique incidence to the sample with a rotating quarter wave plate. The data points (red, blue circles) were fitted to the general model[31–34] (red, blue line):

$$I_{ph}(\alpha)=I_c sin(2\alpha)+I_{L_1}\sin(4\alpha)+I_{L_2}cos(4\alpha)+I_0 \qquad (2)$$

at temperatures of 300 K and 80 K, as in Figure 3a, b, respectively. The oscillatory polarization dependence observed in Figures 3a, 3b originates from the continuous evolution of the incident polarization state as the quarter-wave plate is rotated. Consequently, the measured photocurrent contains contributions from both the circular photogalvanic effect (CPGE) and the linear photogalvanic effect (LPGE). In Eq. (2), the sin(2θ) term represents the helicity-dependent CPGE associated with spin-selective optical excitation, whereas the sin(4θ) and cos(4θ) terms describe the LPGE arising from anisotropic optical absorption and crystal-symmetry-dependent carrier dynamics. The superposition of these contributions naturally gives rise to the observed oscillatory photocurrent response. The sharp phase transition at ~150 K is supported by the temperature resolved photogalvanic measurement (Supporting Figure S4). Figure 3c confirms that the measured circular polarization component of the photocurrent ($I_c$), corresponding to nonlinear injection current,[32,33] is associated here with spin-polarized currents and is positive for any applied voltage and temperature (Supporting Figure S5 and Supporting Table S1-S3), while the linear component (associated with nonlinear shift current) corresponding to linear polarization ($I_{L_2}$) changes sign close to the Néel temperature (see Figure S6). Finally, a small linear component, $I_{L_1}$, is at least an order of magnitude smaller and is neglected; $I_0$ is the polarization independent photocurrent. In our device the presence of a built-in electric field maintains the lack of mirror symmetry and therefore the injection current along the $\hat{z}$ direction (normal to the device plane) does not vanish under the normal illumination conditions of Figure 2. The observed resonant spin polarized photocurrents result from the electronic structure of the heterostructure, under strong reverse bias. Some of these resonances are considered as spectrally protected, as they are robust through the span of entire temperature range, including the magnetic phase transition of $NiPS_3$, as demonstrated in Figure 3d. Additionally, the bias-dependent measurements (Supporting Information, Figures S2, S5, and S6) show that increasing reverse bias enhances the photocurrent magnitude while preserving the characteristic helicity-dependent spectral response, indicating that the reverse bias primarily improves carrier collection efficiency without altering the underlying physical mechanism.

### 2.3 Interfacial electronic structure under an applied electric field

The electronic band structure of a heterostructure comprising three layers of each material was calculated under an electric field of 0.1 $V/Å$ (see more details in the Methods section). Under the electric field, the $d$-states of $NiPS_3$ are shifted towards the middle of the $WSe_2$ bandgap as displayed in Supporting Figure S7. The calculated absorption spectrum (Figure 4a and inset) shows an onset of the absorption at ~1.4 eV and an additional doublet peak at ~1.55 eV, followed by a large band after the ~2 eV peak. The spectral features qualitatively match the photoresponse of Figure 1d, and 2a. The circular dichroism of the absorption (see Methods section) was calculated and integrated on half the Brillouin zone ($0 \leq k_y < 1/2$, Figure 4b), yielding a measure to the photocarrier generation, associated with the measured dichroism, $\eta(\varepsilon, T)$. The band structure along the high symmetry $\Gamma$-$Y$ line of the supercell Brillouin zone is displayed in Figure 4c, with a projection over the $WSe_2$ layers and onto the $\hat{S}_z$ spin state. In this heterostructure, spin-polarized photocurrents exhibit spectral protection, whose microscopic origin lies in the two-step nature of photoconduction: optical absorption followed by directional carrier transport. The interfacial electric field induces a Rashba–Dresselhaus spin–orbit splitting that displaces the two spin branches oppositely in momentum space (Figure 4c). Since the heterostructure preserves global time-reversal symmetry, optical absorption alone is symmetric: each transition at momentum +k into a given spin state is matched by an equivalent transition energy at −k into the opposite spin state, and therefore the absorption by itself enforces no spectral exclusivity on the net spin current. The photocurrent, however, is determined by both absorption and conduction. The direction of current flow, set by the built-in and applied electric fields, selects one half of momentum space, and within this half-space a given photon energy addresses only a single spin branch of the split bands.[35,36] Each excitation energy is thereby locked to a definite spin orientation of the collected carriers, while the opposite spin at the same energy is accessible only upon reversal of the current direction, as dictated by time-reversal symmetry (Supporting Information, Figure S12).[37–39]

Because both ingredients of this locking, the spin–orbit splitting that is of structural and electrostatic origin, and the momentum half-space selection that is of transport origin, are even under time reversal, the resulting selection rules involve no magnetic order parameter. The zigzag antiferromagnetic order of $NiPS_3$, which preserves the combined time-reversal–translation symmetry, therefore leaves the circular photogalvanic response intact; it couples instead to the linear response channels through the magnetic-order-induced lattice anisotropy, consistent with the linear dichroism reported at the Néel transition of zigzag antiferromagnets. [16,40] Moreover, the spin–orbit splitting of the interfacial bands (~$10^2$ meV, Figure 4c) far exceeds the thermal energy over the entire measured range ($k_B T \leq 26$ meV), rendering the spectral selection rules robust against thermal broadening. These considerations directly account for the observed temperature dependence (Figure 3c): the circular component C evolves smoothly across $T_N = 155$ K, exhibiting no sign change or discontinuity at the transition (C = 0.20 ± 0.02 nA at 80 K and 0.24 ± 0.03 nA at 150 K, equal within the fitting uncertainty, Supporting Table S2), and its gradual increase with temperature follows the polarization-independent photocurrent $I_0$, reflecting thermally activated carrier generation and collection rather than a magnetic-order effect. In contrast, the linear component $L_2$, which is sensitive to the anisotropy and to anisotropic carrier scattering, reverses sign close to $T_N$. As a quantitative figure of merit for the spectral protection, we define the polarization retention ratio $R(\varepsilon) = \eta(\varepsilon, 100\ K)/\eta(\varepsilon, 300\ K)$; the protected band at 2.05 eV retains its polarization sign across $T_N$ (R = 1.30), while the band at 1.65 eV reverses sign (R = 1.56).

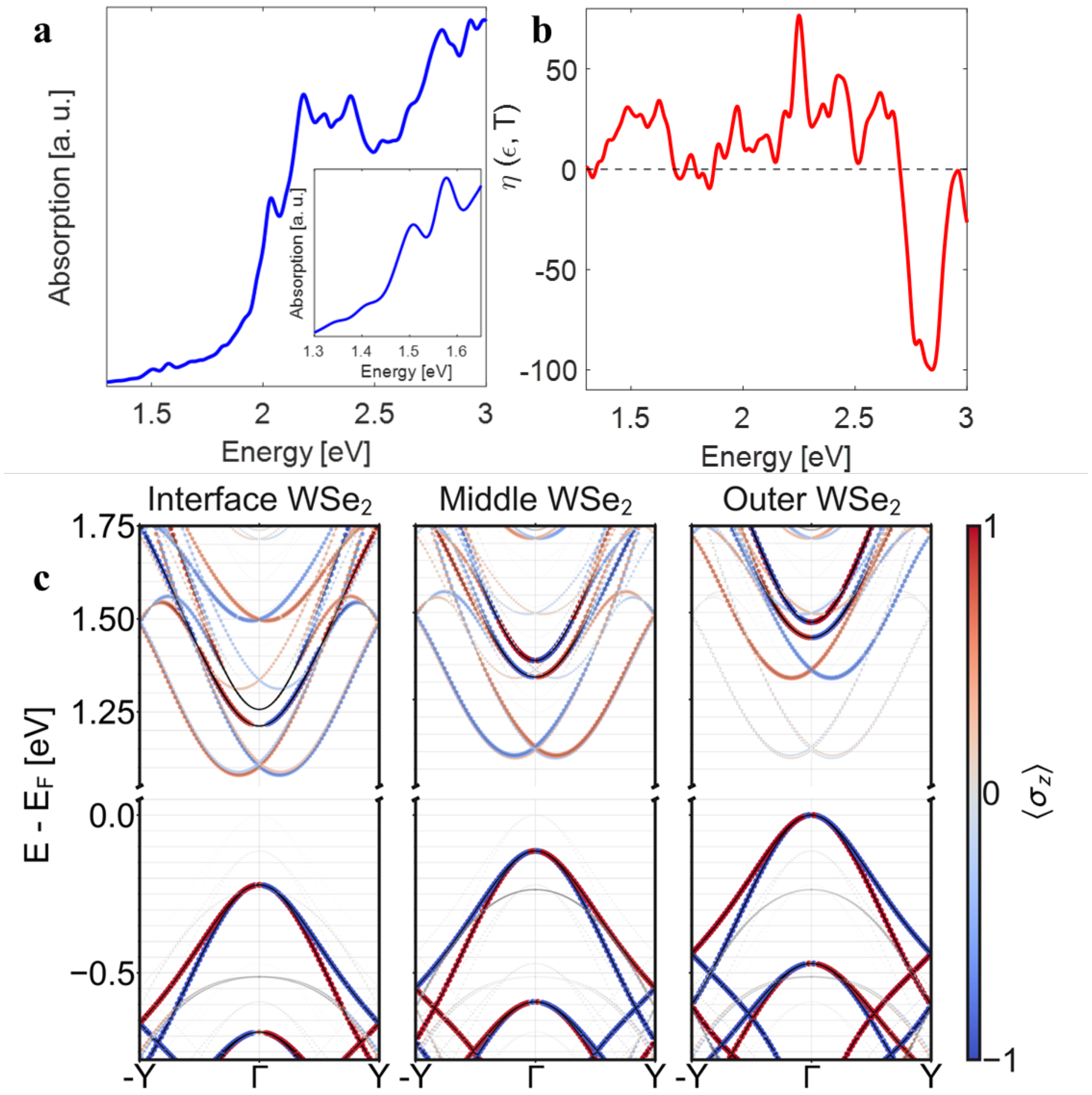

**Figure 4.** Calculated electronic structure of $WSe_2$-$NiPS_3$ interface. a) Calculated absorption spectrum of the heterostructure and b) the corresponding absorption circular dichroism. c) The layer and spin projected band structure of $WSe_2$.

The calculated band structures of Figure 4c reveal a band bending along the electric field (perpendicular to the layers, see also Supporting Figure S8), and RD-like conduction band edges that are key to understanding the underlying mechanism to the observed polarized photocurrents. The shift in k-space of the conduction band valley provides a plausible explanation to our temperature-dependent photocurrent of Figure 2a, c, behaving like a phonon-assisted process. The projected energy bands of Figure 4c reveal that some degree of spin-to-layer locking is taking place, where the spin character changes sign with respect to both the momentum along Y direction and the parity of the layers. The layer-projected optical absorption displayed in Supporting Figure S9, shows for each layer the spectrum of initial and final states of the optical transitions. According to this analysis, the low energy (absorption edge) initial states of the excitation are mostly projected on the $WSe_2$ layers, while the final state is projected on both the $WSe_2$ and the $NiPS_3$ layers. The contribution of $NiPS_3$ to the overall absorption is associated with d-d transitions that are known to be phonon assisted.[19] From the spectral responsivity (Figure 1d and 2a) the onset of the photocurrent appears at ~1.4 eV and interlayer excitations that correspond to shifted band edges at the $NiPS_3$-$WSe_2$ interface are not observed. Since interlayer excitons in 2D materials are known to be lower in amplitude by ~two orders of magnitude,[41] their contribution to the photocurrent in our measurements is below detection threshold.

The calculated electronic structure shows that, despite the preservation of global time-reversal symmetry, the layer- and spin-resolved character of the energy bands originates from local, layer-dependent symmetry breaking.[42] The circularly polarized photoconductive spectral features in Figure 2b, c (see also Figure 1d) arise from this local symmetry: the electric-field enhancement and associated charge transfer are concentrated at the interfacial layers, while contributions from deeper layers are progressively screened. As a result, the system exhibits spectrally protected spin-polarized current generation, manifested by polarization resonances whose sign remains monotonic across the phase transition (Figure 2d and 3c). In this regime, spin-current generation can be controlled through spectral tuning of the excitation rather than by adjusting the polarization state of the incident light.

## 3. Conclusion

To conclude, we have demonstrated a resonant opto-spintronic functionality in a $WSe_2$–$NiPS_3$ vdW heterostructure, where spin-polarized photocurrents are generated and controlled through spectral selectivity rather than external magnetic fields or polarization modulation at 80% efficiency close to the magnetic transition temperature and 30% efficiency at room temperature. The observation of complementary circularly polarized photoconductive resonances, including spectrally protected bands whose polarization remains invariant across the antiferromagnetic phase transition of $NiPS_3$, reveals a robust mechanism for spin-current generation rooted in local symmetry breaking at the heterointerface. Photogalvanic measurements identify a dominant injection-current contribution associated with circular polarization, while first-principles calculations show that electric-field-induced interfacial hybridization and layer-localized states give rise to spin–layer locking despite preserved global time-reversal symmetry. These findings establish spectral tuning as an effective control knob for spin transport in antiferromagnetic heterostructures and position correlated van der Waals magnets as a versatile platform for resonant opto-spintronics, with prospects for spectrally programmable spin sources and integrated quantum opto-spintronic devices. The heterostructure interfaces and their corresponding excitations can be dynamically tuned by strain, electric fields, and temperature. The temporal response of optical spin injection is governed primarily by the optical relaxation time and therefore holds promise for ultrafast control. Together, these aspects provide fertile ground for future studies.

## 4. Experimental Section

**4.1 Device fabrication:** The $WSe_2$-$NiPS_3$ heterostructure was fabricated using a PDMS-assisted dry transfer method performed within a nitrogen-filled glovebox to prevent contamination and degradation. High-quality hexagonal boron nitride (hBN) and tungsten diselenide ($WSe_2$) crystals were purchased from HQ Graphene, while bulk nickel phosphorus trisulfide ($NiPS_3$) crystals were synthesized in-house via the chemical vapor transport (CVT) technique[43]. A patterned back electrode (Ti/Au, 3/27 nm) was defined on 285 nm $Si/SiO_2$ substrates in liftoff process. Thin flakes of $WSe_2$ and $NiPS_3$ were mechanically exfoliated from their respective bulk crystals onto PDMS stamps. Using the dry transfer method, $NiPS_3$ flakes were first transferred onto the pre-patterned back electrode, followed by the placement of $WSe_2$ flakes directly atop the $NiPS_3$ layer. The resulting heterostructure was then encapsulated with an hBN flake.

**4.2 Cross-Sectional TEM Sample Preparation and Imaging:** Cross-sectional lamellae of the $WSe_2$-$NiPS_3$ were prepared using a focused ion beam (FIB) milling technique. Imaging was conducted using a probe-corrected FEI Titan G2 80–200 ChemiSTEM TEM operating at an accelerating voltage of 200 kV.

**4.3 Photoconductivity Measurements:** Monochromatic and spectrally resolved photocurrent measurements were conducted under vacuum (~$10^{-5}$ Torr) across a controlled temperature range of 80–300 K. Current–voltage (I–V) characteristics were obtained using a precision source-measure unit (Keysight B2900). The devices were mounted in a temperature-controlled Linkam stage (HFS350E-PB4) with a quartz window for optical access. For spectral photocurrent response, the devices were electrically interfaced with a Thermo Fisher Scientific Nicolet iS50R Fourier Transform Infrared (FTIR) spectrometer, coupled to a Nicolet Continuμm microscope using a Quartz–Halogen light source and a Quartz beam splitter.

**4.4 Circular photo-galvanic measurements:** A 632.8 nm He-Ne laser served as the light source for circular photo galvanic effect (CPGE) measurements. Here the laser beam was significantly larger than the device active area illuminated the sample at oblique incidence of 45º. Control over polarized light was achieved by passing linearly polarized laser light through a rotatable quarter-wave plate. The modulated light beam illuminated the sample, and the resulting photocurrent was detected through two contacts, amplified using a low-noise current preamplifier, and a SR830 lock-in amplifier synchronized to an optical chopper at 1 kHz.

**4.5 Ab-initio calculations:** The model of the trilayer-$WSe_2$–trilayer-$NiPS_3$ heterostructure was created by using the corresponding bulk-type stackings and by employing an in-plane $\begin{pmatrix} 3 & 3 \\ -1 & 1 \end{pmatrix}$ supercell of $WSe_2$ and the magnetic unit cell of $NiPS_3$. The interlayer twist was 60°, i.e., the hexagons of $WSe_2$ are rotated by 60° with respect to the hexagonal pattern of Ni atoms. We checked with fewer layers that other twists – which leads to much larger supercells – have a negligible influence on the relative alignment of the bands. The resulting structure contains 114 atoms. The lattice was fixed to match the bulk lattice parameter of $WSe_2$ which leads to a compressive strain on $NiPS_3$ of -1.5%. To fully relax the atomic positions, we used FHI-aims[44] employing the Perdew-Burke-Ernzerhof[45] functional on intermediate tier 1 numeric atom-centered orbitals, including the nonlocal many-body dispersion correction,[46,47] and scalar relativistic corrections on a 10×6×1 Γ-centered **k**-grid. The electronic band structure, the Mulliken projections, and the density of states were calculated including spin-orbit coupling[48] and considering the dipole correction on a 20×12×1 Γ-centered **k**-grid and furthermore using the SCAN functional.[49] For the calculation of the momentum-matrix elements (MMEs) we increased the **k**-point sampling to 32×20×1 and furthermore determined the band dispersion in the full Brillouin zone (BZ) using 5808 **k** points with a finer mesh close to the Γ point (backfolded K point of $WSe_2$). Since the absolute square of the MMEs is proportional to the oscillator strength of a transition,[50] we estimated the absorption for different polarizations of the light by integrating the MMEs over the full/half BZ including 60 valence and 78 conduction bands and by including a convolution with a Gaussian function with a root-mean-square width of 0.02 eV. Using this approach neglects the local-field effects (random-phase approximation). We furthermore calculated a layer-projected absorption by using the MMEs and the Mulliken projections of the initial and final states along a path through the BZ. As this only takes into account a small amount of all possible excitations, this can however only give a qualitative picture.

**4.6 Statistical Analysis.** The present study is based on experimental photocurrent measurements and first-principles calculations performed on a representative $WSe_2$-$NiPS_3$ heterostructure devices. Helicity-dependent photocurrent polarization (η) was calculated directly from the measured photocurrents according to Eq. (1) and therefore does not involve statistical fitting. For polarization-dependent photocurrent measurements, the uncertainties of the fitted parameters were obtained from the fitting procedure and are presented as error bars where applicable. Statistical hypothesis testing, P values, and sample-size comparisons are not applicable to the present study.

DATA AVAILIBILITY

Data available on request from the authors.

ASSOCIATED CONTENT

Supporting Information available: Supporting text, figures and tables.

AUTHOR INFORMATION

* Corresponding author e-mail: Doron.Naveh.biu.ac.il, ssefrat@technion.ac.il, thomas.heine@tu-dresden.de

## COMPETING INTERESTS

The authors declare no competing interest.

## ACKNOWLEDGMENTS

All authors would like to thank the Deutsch – Israel Program (DIP) for supporting this work with grant No. NA1223/2-1. We gratefully acknowledge the computing time made available to them on the high-performance computer at the NHR Center of TU Dresden and on the high-performance computers Noctua 2 at the NHR Center PC2. These are funded by the German Federal Ministry of Education and Research and the state governments participating based on the resolutions of the GWK for the national high-performance computing at universities (www.nhrverein.de/unsere-partner). N. C. acknowledges the Alexander von Humboldt foundation for a Postdoctoral Fellowship.

## AUTHOR CONTRIBUTION

AH, NC and EL synthesized the crystal samples, RKY, MP, AL and DN fabricated and measured devices, TB and TH first principles calculations and theoretical analysis. All authors contributed to the manuscript writing.